\documentclass[pra,aps,twocolumn,showpacs]{revtex4-1}

\usepackage[english]{babel}
\usepackage{multirow}
\usepackage{pifont}                           
\usepackage{amssymb}                           
\usepackage{amsmath}                           
\usepackage{graphicx}                          
\usepackage{bm}                                
\usepackage{color}
\usepackage{psfrag}
\usepackage{fix-cm}

\begin{document}

\title{Phase coherence in
quasicondensate experiments: an \textit{ab initio} analysis via the stochastic Gross-Pitaevskii equation}

\author{D.\ Gallucci}
\author{S.\ P.\ Cockburn}
\author{N.\ P.\ Proukakis}

\affiliation{Joint Quantum Centre (JQC) Durham-Newcastle, School of Mathematics and Statistics,\\ Newcastle University, Newcastle upon Tyne, 
NE1 7RU, United Kingdom
}



\begin{abstract}

We perform an {\em ab initio} analysis of the temperature dependence of the 
phase coherence length of finite temperature, quasi-one-dimensional Bose gases
measured in the experiments of Richard {\it et al}. (Phys. Rev. Lett. 
\textbf{91}, 010405 (2003)) and 
Hugbart {\it et al.} (Eur. Phys. J. D \textbf{35}, 155-163 (2005)),
finding very good agreement across the
entire observed temperature range ($0.8<T/T_{\phi}<28$).
Our analysis is based on the one-dimensional stochastic Gross-Pitaevskii equation, modified to self-consistently
account for transverse, quasi-one-dimensional effects, thus making it a valid model in the regime $\mu\sim {\rm few}
\hspace{0.1cm} \hbar\omega_\perp$. 
We also numerically implement an alternative identification of $T_{\phi}$, based on direct analysis of the distribution of phases in a stochastic treatment.

\end{abstract}

\maketitle

\section{Introduction}

Weakly interacting ultracold Bose gases are of great interest because they allow the `pure' study of quantum phenomena on a macroscopic scale. 
Such systems are typically engineered in harmonic traps with different geometric configurations, whose dimensionality plays a crucial role in determining the properties of these gases.
In three-dimensional harmonic traps, the system undergoes a phase transition which leads to the appearance of coherence across the entire sample, as demonstrated in \citep{Andrews1997,Hagley1999,Stenger1999,Bloch2000}.
By setting the trap frequency in one direction to be much larger than the others, the effective system dynamics is reduced to two dimensions \citep{Gorlitz2001,Rychtarik2004,Stock2005,Smith2005}, with many interesting phenomena occurring, such as the Berezinskii-Kosterlitz-Thouless transition \citep{Hadzibabic2006,Schweikhard2007,Kruger2007,Hadzibabic2008,Clade2009,Rath2010,Tung2010,Hung2011}. Increasing the trap frequency in a further direction allows the realization of highly elongated traps, where the interesting physics occurs in the axial direction \citep{Petrov2001,Gorlitz2001,Schreck2001,Dettmer2001,Shvarchuck2002,Hellweg2003,Richard2003,Moritz2003}.

In such a regime two characteristic temperatures become relevant, associated with the onset of phase ($T_{\phi}$) and density ($T_{\rm d}$) fluctuations \cite{Petrov2000}. For temperatures $T_{\phi}<T<T_{\rm d}$, density fluctuations tend to be suppressed, and the system reduces to a condensate with fluctuating phase (quasicondensate) \cite{Popovbook}.
In such a geometry low-energy thermal excitations of the axial modes play a crucial role, as they tend to destroy the coherence in the sample \citep{Popov1972,Popovbook,Petrov2000,Andersen2002,AlKhawaja2002,AlKhawaja2003,Proukakis2006c,Luxat2003,Mora2003,Bogoliubov2004,Gerbier2004,Kadio2005,Gerbier2003,Hellweg2001,Cacciapuoti2003,Bouyer2004,Hugbart2005}; such excitations may have wavelengths greater than the transverse extent of the system, therefore acquiring a one-dimensional (1D) character \cite{Hellweg2001,Dettmer2001,Shvarchuck2002,Hellweg2003,Cacciapuoti2003,Gerbier2003,Richard2003,Bouyer2004,Hugbart2005,Schumm2005,Trebbia2006,Esteve2006,Hofferberth2007,vanAmerongen2008,Manz2010,Betz2010,Kruger2010,Armijo2010}.  
An accurate analysis of the coherence properties in such systems is therefore necessary for potential applications, such as matter-wave interferometry \citep{Shin2004,Hinds2001,Hansel2001,Andersson2002,Schumm2005,Wang2005,Jo2007}, atom chips \citep{reichel_vuletic_book_11} and atom lasers \citep{Bloch1999,Hagley1999,Gerbier2001,Lahaye2004,Guerin2006}. 
Although experiments can nowadays be engineered to produce gases which are both weakly interacting and
 practically 1D ($\mu,k_{B}T  \lesssim \hbar\omega_\perp$, where $\mu$ is the
 chemical potential, $k_{B}T$ the thermal energy and $\hbar\omega_\perp$ the transverse excitation energy) \citep{Moritz2003,Schumm2005,Trebbia2006,Esteve2006,Hofferberth2007,vanAmerongen2008,Armijo2010,Kruger2010,Manz2010,Betz2010}, the early 
experiments performed did not satisfy these conditions so well, and the system was instead in the 1D-3D crossover regime \citep{Petrov2001,Dettmer2001,Shvarchuck2002,Richard2003,Gerbier2003,Hellweg2003,Hellweg2001,Cacciapuoti2003,Bouyer2004,Hugbart2005}. 

In this work we propose and implement a modified stochastic model which enables us to perform a successful \textit{ab initio} description in the `intermediate' regime $\mu, k_B T \sim {\rm few} \,\, \hbar \omega_\perp$ where quasi-condensate physics dominates, but transverse effects still need to be appropriately accounted for. More specifically, we analyze two early experiments \citep{Hugbart2005,Richard2003} investigating the phase coherence properties of weakly interacting, quasi-1D Bose gases.
These experiments measured the temperature dependence of the coherence length $L_{\rm c}$, over which the gas maintains an appreciable coherence, both in the `strong' ($6<T/T_{\phi}<28$) \cite{Richard2003} and `weak' ($0.8<T/T_{\phi}<8$) \cite{Hugbart2005} phase fluctuation regimes.
Although it is theoretically anticipated (for a homogeneous gas \citep{Cazalilla2004}), that the coherence length, scaled to the experimental half-length of the system $L$, 
should yield a universal curve when plotted against the reduced temperature $T/T_{\phi}$, it would not actually be appropriate to incorporate their {\em reported} data
into a single graph \cite{priv_comm} 
spanning the entire regime $0.8 < T/T_{\phi} < 28$, because the two experiments used different techniques to extract the coherence.

The main achievements of this work are as follows:
(i) we 
show that the measurements reported in both experiments are indeed {\em consistent} with such a unified curve; this is achieved by a completely {\em ab initio} analysis of the experimental data, using only the quoted experimental values for atom number, trap frequencies and temperature; (ii) our analysis, which is based on the quasi-1d stochastic Gross-Pitaevskii equation \cite{Cockburn2011a}, appropriately modified here to extend its validity to the regime $\mu \sim {\rm few} \, \hbar \omega_{\perp}$ reveals excellent agreement in the `strong' phase fluctuation regime; (iii) our attempt to more closely mimick the experimental procedure of \cite{Hugbart2005} -- whose findings have not been adequately interpreted to date -- also yields good overall agreement (within error bars); in order to further improve this we implement an alternative approach of extracting a phase coherence temperature based on direct analysis of the phase distributions from different realisations of our stochastic treatment.

\section{Methodology}

Our treatment is based on a modified one-dimensional form  \citep{Cockburn2011a} of the stochastic Gross--Pitaevskii equation (SGPE) \citep{Stoof1999,Stoof2001,Gardiner2003}.
The modification proposed in \citep{Cockburn2011a} -- which was found to be essential  
to accurately simultaneously reproduce both {\it in situ} density profiles and density fluctuations of recent quasi-one-dimensional 
 experiments \citep{Trebbia2006,vanAmerongen2008,Armijo2010} -- takes the form: 
\begin{eqnarray}  
i\hbar\frac{\partial \psi(z,t)}{\partial t}=[1-i\gamma(z,t)]\bigg(-\frac{\hbar^2}{2m}\frac{\partial^2}{\partial z^2}+ V(z) \nonumber\\
+\hbar\omega_\perp\big[\sqrt{1+4|\psi|^2a_s}-1\big] -\mu \bigg)\psi(z,t)+\eta(z,t)\label{SGPE}.
\end{eqnarray}
Here $\psi$ describes the relevant (highly populated) low-energy modes of the system, $V(z)=m \omega_{z}^{2} z^2/2$ is the axial trapping potential and $\eta$ is a complex Gaussian noise term, with correlations $\langle\eta^*(z,t)\eta(z',t')\rangle=2\hbar\gamma(z,t)k_BT\delta(z-z')\delta(t-t')$, where $\gamma(z,t)$ denotes the dissipation. (For further details on numerical implementations and related treatments see 
Refs.~\citep{Cockburn2011a,Stoof1999,Stoof2001,Stoof1998,Duine2001,AlKhawaja2002,Proukakis2003_lasphys,Proukakis2006a,Cockburn2009,Cockburn2010,cockburn_nistazakis_11,Cockburn2011b,Gardiner2003,Weiler2008,Bradley2008,Rooney2010,Damski2010,Proukakis2008,Blakie2008}). 
The condition $\mu \lesssim {\rm few}\hspace{0.1cm} \hbar\omega_{\perp}$ can lead to a swelling of the condensate in the transverse direction, relative to the true 1D transverse ground state; this quasi-1D effect is due to repulsive interactions and is reproduced by using the modified nonlinear term $\hbar\omega_\perp\big[\sqrt{1+4|\psi|^2a_s}-1\big]$ (where $a_s$ is the s-wave scattering length), which reduces to the 1D result in the limit $4|\psi|^2 a_s\ll1$.
For the ordinary Gross-Pitaevskii equation, this was shown 
in Refs.~\citep{Salasnich2002,Fuchs2003,Gerbier2004,Mateo2007,Frantzeskakis2010}, and its validity was verified experimentally in \cite{Kruger2010}.

In order to match to experimental atom numbers, it is crucial to also include in our treatment the contribution ($n_\perp$) to the linear density profile
of transverse thermal atoms with energy greater than $\hbar\omega_{\perp}$.
This is done via
\begin{eqnarray}
n(z;\mu,T)=\langle|\psi(z;\mu,T)|^2\rangle + n_\perp(z;\mu,T)
\label{tot_dens}
\end{eqnarray}
where $\langle\cdots\rangle$ denotes ensemble averaging, obtained by averaging over many different realizations of the noise, 
and
\begin{eqnarray}  
n_\perp(z;\mu,T)=\frac{1}{\lambda_{dB}}\sum_{j=1}^{\infty}(j+1)g_{1/2}\left(e^{\mu_j(z)/k_{B}T}\right)
\label{eq:nperp}
\end{eqnarray}
where $g_{1/2}(...)$ is the polylogarithm (or Bose function) of order $1/2$, and $\lambda_{dB}=h/\sqrt{2\pi m k_BT}$ is the thermal de Broglie wavelength
\cite{therm_atoms}. While in the limit $\mu<\hbar\omega_\perp$ studied in our previous work \citep{Cockburn2011a}, it is sufficiently accurate to use $\mu_j(z)=\mu-V(z)-j\hbar\omega_\perp$ in Eq.~(\ref{eq:nperp}) (see also \citep{vanAmerongen2008,Davis2011}),
the effect of mean-field potential experienced by the transverse thermal atoms should also be taken into account in the regime $\mu>\hbar\omega_\perp$, characteristic of the experiments of Refs.~\cite{Richard2003,Hugbart2005} studied here. For this reason we use here the modified expression:
\begin{equation}
\mu_j(z)=\mu-V(z)-j\hbar\omega_\perp-2g(\langle|\psi|^2\rangle+n_\perp)
\label{mu_j}
\end{equation}
where $g = 2 a_s \hbar \omega_\perp$ is the effective interaction strength.

In the following sections we systematically use Eqs.~(\ref{SGPE})--(\ref{mu_j}) to match to the experimental atom number (noting also that the profiles generated by Eq.~(\ref{tot_dens}) correspond to transversely {\em integrated} density profiles, typical of ultracold gas experiments), before performing an analysis of the coherence properties.

For both experiments analysed here, the {\em only} experimental inputs to our theory are the atomic species used ($^{87}$Rb in both cases), the trapping configuration, the temperature and atom number corresponding to {\em each reported experimental data point} \cite{Jocelyn}. 
In our {\em ab initio} analysis we fix experimental trap configuration and temperature, and vary $\mu$ until the desired number of atoms is reached in our simulations. Depending on the available experimental data, we either match to the {\em total} or the {\em quasicondensate} experimental atom number, as explained in each case. Representative error bars are also calculated for our simulations, based on variations in both number and temperature, in close analogy to the experimental analysis.

In order to discuss both experiments in a coordinated way, and demonstrate their consistency, we consider the dependence of the coherence length on temperature, with both parameters appropriately scaled to show the emergence of universal physics. In performing this analysis, we also address experiment-specific features, meaning the precise definitions of the coherence length and phase fluctuation temperature differ slightly within our subsequent analysis, however this difference is clearly labelled within figures.

\section{Comparison to the experiment of Richard et al. \cite{Richard2003} \label{exp_strong}}

The first experiment we consider is that of Richard $et$ $al.$ \cite{Richard2003}, in which a very elongated harmonic trap ($\omega_\perp/2\pi=760$ Hz and $\omega_z/2\pi=5$ Hz) was used to generate quasicondensates of atom  numbers in the range $0.25\times10^5-0.65\times10^5$ and at temperatures $90$ nK $ \le T \le 350$ nK.
These parameters enabled the `strong' phase fluctuation regime $6<T/T_{\phi}<28$ to be probed.

In order to investigate the phase coherence properties of the gas, the axial momentum distribution was measured by means of Bragg spectroscopy
and found to have a Lorentzian shape, consistent with an exponential decay of the correlation function in space anticipated in the regime of large phase fluctuations \citep{Richard2003,Hellweg2003,Cacciapuoti2003,Gerbier2003}.
This approach enabled the coherence length $L_{\rm c}$ to be extracted from the half width at half maximum of the momentum profile, and its temperature dependence to be measured.
Using the original data relevant for Ref.~\cite{Richard2003} (see also \cite{Bouyer2004} and \cite{parameters}), we can plot this in universal form by scaling the
coherence length $L_{\rm c}$ to the half-length of the condensate $L$, and plotting against the reduced temperature $T/T_{\phi}$. 
Here we use the general definition of $T_{\phi}$ in terms of the 1D axial quasicondensate peak density, $n_{{\rm qc}}(0)$, i.e.\ \cite{Petrov2001,Gerbier2004}
\begin{eqnarray}
T_{\phi}[n_{{\rm qc}}(0)]=\hbar^2 n_{{\rm qc}}(0)/m k_B L \;,
\label{T_phi_nqc}
\end{eqnarray} 
which is relevant for Refs.~\cite{Richard2003,Bouyer2004}.
(See discussion following Eq.~(7) for a closely related expression valid in the Thomas-Fermi regime and used in the analysis of Ref.~\cite{Hugbart2005}.)

\begin{figure}
  \includegraphics[scale=0.33]{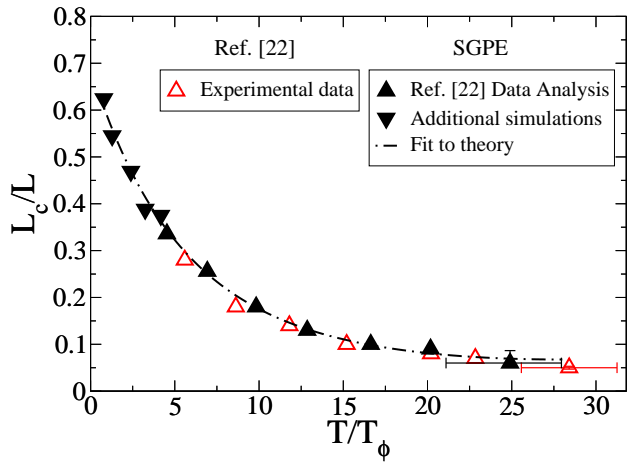}
   \caption {(Color online) Coherence length $L_{\rm c}$ scaled to the half length of the quasicondensate $L$ vs. reduced temperature $T/T_{\phi}$. Comparison of experimental data \cite{Richard2003,Bouyer2004} spanning the range $T/T_{\phi} > 6$ (hollow red triangles) to the SGPE model (upward filled black triangles). Additional numerical results generated with the SGPE model for lower temperatures $T/T_{\phi} < 6$  are also shown (downward filled black triangles), together with the best fit of the SGPE results (dot-dashed black line). 
The horizontal experimental error bar is taken as 10\% \cite{Richard2003}, based on typical experimental uncertainties,
while the corresponding error in the simulated point arises from a typical $15\%$ variation in the quasicondensate atom number and 5\% variation in temperature \cite{Richard2003}; in both cases the vertical error bars fall within the symbol size.
Temperatures have been scaled to $T_{\phi}[n_{{\rm qc}}(0)]=\hbar^2 n_{{\rm qc}}(0)/m k_B L$, where $n_{{\rm qc}}(0)$ is the peak quasicondensate density; in the SGPE analysis, the spatial extent of the quasicondensate is obtained \textit{ab initio} as the temperature dependent Thomas-Fermi radius $R_{\rm TF}(T)$ \label{strong_phase}\cite{R_TF}. }
\end{figure} 

In order to simulate the experiment, we compute the first order correlation function in position space $g^{(1)}(-z/2,z/2)=\langle\psi^*(-z/2)\psi(z/2)\rangle$ (normalized to the averaged central density $\langle|\psi(0)|^2\rangle$), finding an exponential behaviour against distance $z$ as in the experiment
(since $T \gg T_{\phi}$ here).
We then compute the Fourier transform $C^{(1)}=F[g^{(1)}]$ and extract the coherence length $L_{\rm c}$ by measuring $\Delta k$, the half-width at half maximum of $C^{(1)}$.
We also use our simulated results to find $T_{\phi}[n_{{\rm qc}}(0)]$, with 
the required inputs being the quasicondensate peak density and spatial extent of the gas, each of which we obtain \textit{ab initio}. Specifically, 
the quasicondensate density is extracted as \cite{Svistunov2001,Proukakis2006b,bisset2009a,Cockburn2011b}: 
\begin{equation}
n_{{\rm qc}}(z)=\sqrt{ 2\langle|\psi(z)|^2\rangle^2-\langle|\psi(z)|^4\rangle}
\label{quasicondensate}
\end{equation}
and its peak value is used in the expression for $T_{\phi}[n_{{\rm qc}}(0)]$.
Moreover, the half-length of the quasicondensate in our simulations is given by a temperature dependent Thomas-Fermi radius $R_{\rm TF}(T)$, 
extracted from Eq.~(\ref{SGPE}) analogously to the procedure used in the modified Popov theory (see \citep{Andersen2002,AlKhawaja2002,Proukakis2006b} and \cite{R_TF}).

Our {\em ab initio} SGPE results presented in Figure~\ref{strong_phase} (upward filled black triangles) show excellent agreement to the experimental results (hollow red triangles) in this `strong' phase fluctuation regime within their respective characteristic error bars (see caption for details). 
Note that the coherence length is smaller than the quasicondensate extent ($L_{\rm c}/L<1$) in the temperature range probed here, illustrating the fundamental role of phase fluctuations in such an elongated geometry.
As $T/T_{\phi}$ approaches zero, we expect the coherence length to increase: to show this, we have generated a further set of numerical points for $T/T_{\phi}<6$ (downward filled black triangles), which indeed confirm this picture, and the `universal nature' of such a scaled diagram. 

In the following section, we turn to the investigation of the experiment of Hugbart et al.~\cite{Hugbart2005}, in which the regime $T/T_{\phi}<8$ was probed using different methods.

\section{Comparison to the experiment of Hugbart et al. \cite{Hugbart2005}}

The Bragg spectroscopy method used in the previous experiment limited accurate investigations of the coherence properties of the gas to $T/T_{\phi}>6$, as at lower
temperatures the width of the momentum distribution of the gas was no longer easily resolved \cite{Bouyer2004}.
In the experiment described in \cite{Hugbart2005}, an alternative interferometry technique (see also Ref.~\cite{Hellweg2003}) was used to measure the spatial correlation function in the regime $0.8<T/T_{\phi}<8$.

\begin{figure}
  \includegraphics[scale=0.3]{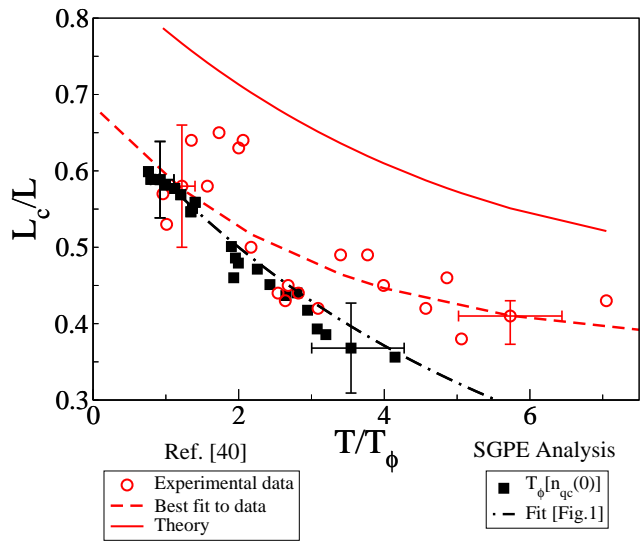}
   \caption {(Color online) Scaled coherence length $L_{\rm c}/L$ vs. reduced temperature $T/T_\phi$. Comparison of experimental data \cite{Hugbart2005} (hollow red circles) to predictions of the SGPE model (filled black squares). The reported experimental best fit (dashed red line) and the theory used in \cite{Hugbart2005} to interpret the experimental data (solid line) are also shown, alongside the SGPE fit from Fig.~\ref{strong_phase} (dot-dashed black line). Typical error bars for corresponding data points in both experiment and SGPE model are indicated, with the SGPE ones arising from a characteristic 20\% variation in total atom number and an additional 5\% variation in temperature;
(note that the SGPE error bars for the point with $T/T_{\phi} \approx 1.2$ lie within the point size and are not visible). The SGPE results are scaled to $T_{\phi}[n_{{\rm qc}}(0)]$ for consistency with the discussion of Fig.~\ref{strong_phase}. \label{strongandweak} }
\end{figure}

In particular, after the condensate was released, two Bragg pulses were applied, playing the role of matter-wave beam splitters.
The contrast of the resulting interference fringes was then obtained from the modulus of the Fourier transform of the interference pattern, and the coherence length was extracted from the decrease of the contrast as a function of the distance between the two interfering condensates. 
The experimental data (hollow red circles of Fig.~\ref{strongandweak}), show that, in the regime $T\simeq T_{\phi}$, the coherence extends over more than half of the system size. 

In order to access such low values of $T/T_{\phi}$, it was technically easier to use slightly {\em less} elongated traps than in \cite{Richard2003}. In fact, two different trap configurations were used 
(first: $\omega_\perp/2\pi=395$ Hz and $\omega_z/2\pi=8.67$ Hz; second: $\omega_\perp/2\pi=655$ Hz and $\omega_z/2\pi=6.55$ Hz); the data obtained with the second trap were subdivided into two different blocks, characterised by two different values of the evaporation parameter, which proved necessary in order to vary $T/T_{\phi}$, while keeping the condensed fraction fairly constant within each data block. The {\em total} atom numbers measured in this experiment were found to lie in the 
range $0.8\times10^5-3\times10^5$ (corresponding quasicondensate \cite{cond_terminology} atom numbers: $0.5\times10^5-2.5\times10^5$), within a temperature region of $100-230$ nK
\cite{Jocelyn}.

In our simulations, rather than reproducing the experimental procedure, for which non-equilibrium expansion dynamics would need to be accounted for, we {\em initially} instead extract the corresponding {\em in situ} coherence length by adopting the {\em same} methodology described in Section~\ref{exp_strong}.
This is done here in order to firstly 
explore whether our results for this new system also lie on the same `universal' curve as those of Ref.~\cite{Richard2003} (dot-dashed black curve of Fig.~\ref{strong_phase}) when analysed in an identical manner.

We indeed find that the new numerical results generated in this regime $T\sim {\rm few} \, T_{\phi}$ (filled black squares of Fig.~\ref{strongandweak}) lie on the same (dot-dashed black) curve provided by the fit in Fig.~\ref{strong_phase}; this demonstrates the universal character of the coherence properties, as numerical results obtained with different sets of trap configurations, temperatures and atom numbers yield the same behaviour.
As the analysis of Ref.~\cite{Hugbart2005} reported error bars for two indicative data points, we
repeat a similar analysis within the SGPE theory -- in which case it is based on a 20\% variation in
total atom number \cite{Hugbart_PhD}, and a typical 5\% variation in temperature \cite{frac_err,Gerbier2004_pra}.

However, while our methodology led to a unified theoretical graph over both the `weak' and `strong' phase fluctuation regime, which also provides excellent agreement with the experiment \cite{Richard2003} for $T > 6T_{\phi}$, the corresponding SGPE results for the experiment of Ref.~\cite{Hugbart2005} differ from the experimentally-reported ones. The reason for this discrepancy is twofold: firstly, in \cite{Hugbart2005} it is an `effective' correlation function that is evaluated, which leads to a 
different definition of the coherence length \cite{priv_comm} (see Section~\ref{C_eff}); secondly, the theoretical values of $T/T_{\phi}$ obtained in our treatment do not span precisely the same range as in the experiment (see Section~\ref{Ident_Tphi}).   

With regard to the latter point, we note that the analysis of Ref.~\cite{Hugbart2005} was actually based on a {\em slightly} modified definition for $T_{\phi}$ compared to that of Eq.~(5); in particular, in \cite{Hugbart2005}, $T_{\phi}$ was defined in terms of the {\em number} of quasicondensate \cite{cond_terminology} atoms $N_{{\rm qc}}$ (and not on the peak quasicondensate {\em density}), via the expression \cite{L_definition}
\begin{equation}
T_{\phi}[N_{{\rm qc}}]=15\hbar^2 N_{{\rm qc}}/16 m k_B L^2 \;.
\label{T_phi_N}
\end{equation}  
This is in fact a simplified form of Eq.~(5), valid for 3D condensates \cite{reconcile_tphi} 
(i.e.\ condensates where the transverse density profiles can be well approximated by a Thomas-Fermi profile).

In the following sections, we further investigate potential sources of discrepancy between our simulations and the experimental results of Ref.~\cite{Hugbart2005}, focusing our analysis on a modified effective correlation function defined below, as relevant for this experiment.
Throughout our analysis, the coherence length is always referred to as $L_{\rm c}$, with the method of its extraction clearly identified in each figure.

\subsection{Effective correlation function \label{C_eff}}

The main source of the observed discrepancy should be related to the experimental measurement of an `effective' correlation function (see Eq. (9) in \cite{Hugbart2005}), instead of the correlation $C^{(1)}$ measured in Richard {\em et al.} \cite{Richard2003} and discussed thus far. 
In the experiment of Ref.~\cite{Hugbart2005}, this effective correlation was introduced in order to cancel the random phase caused by the shot to shot fluctuations of the global position of the contrast fringes;  
it was found that taking the absolute value of the Fourier transform of the fringe pattern before averaging achieved this aim (for a more detailed explanation see Section 4.2 in Ref.~\cite{Hugbart2005}), but modified the coherence length relative to that of $C^{(1)}$. For this reason, we should not expect the measurements from the two experiments to lie on the same curve, as they measure two different quantities \cite{priv_comm}.

\begin{figure}[h]
  \includegraphics[scale=0.318]{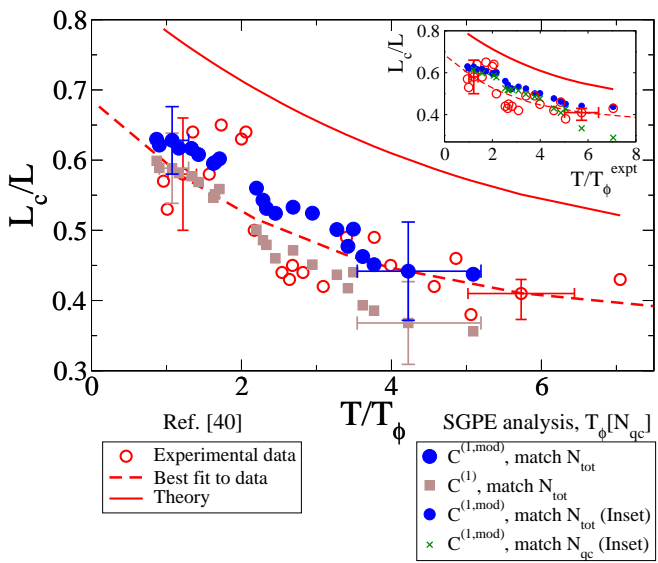}
 \caption{(Color online) Scaled coherence length $L_{\rm c}/L$ vs. reduced temperature $T/T_\phi$. Comparison of numerical SGPE findings extracted via $C^{{(\rm 1,mod)}}$ (filled blue circles) and via $C^{(1)}$ (filled brown squares) against experimental data points of Ref.~\cite{Hugbart2005} (hollow red circles). The theory used in \cite{Hugbart2005} is also shown (solid line) together with the experimental best fit (dashed red line). 
SGPE numerical points are scaled to $T_{\phi}[N_{{\rm qc}}]$, and respective error bars account for $20\%$ variation in total atom number and a further 5\% variation in temperature. 
Inset: comparison of SGPE results, in which temperatures are scaled to the {\em experimentally quoted} values of $T_{\phi}$. 
SGPE results shown in the inset are obtained by matching either the {\em total} number of atoms (blue filled circles, main figure and inset), or the quasicondensate atom number  (green crosses, inset only).
\label{weak} }
\end{figure}

In analogy to the experimental method used to extract the coherence length, we implement this feature by taking the absolute value of the Fourier transform of 
$g^{(1)}$ from {\em each individual run}, before averaging over the different realizations of the noise; our `effective' correlation function, which we here call $C^{{(\rm 1,mod)}}$, takes the form  $C^{{(\rm 1,mod)}}=\langle\big|F[g^{(1)}(-z/2,z/2)]\big|\rangle$.
The `effective' correlation function is found to have similar behaviour to $C^{(1)}$, but it decays faster (in the region of interest) in momentum space, 
thus resulting in larger values of the coherence length. 
This is shown in Fig.~\ref{weak}, which compares the theoretical results to the experimental measurements using $C^{(1)}$ (brown filled squares)
and $C^{(1,{\rm mod})}$ (blue filled circles).
Although we consciously 
do not exactly reproduce the experimental measurement sequence, the results obtained from the SGPE analysis of the `effective' correlation 
function clearly show a very similar trend to the experimental findings over the probed regime; in particular, they tend to lie on the reported line of best fit of the experimental data (dashed red line). 
We note that our calculation of $C^{{(\rm 1,mod)}}$ leads to a much improved agreement with the experimental data than the original theoretical analysis reported in \cite{Hugbart2005} (solid red line)  
which largely overestimates the amount of coherence in the system.

Although the experimental results and our simulated points (extracted via $C^{{(\rm 1,mod)}}$) demonstrate very good agreement when accounting for their respective error bars, the 
experimental data still appear to systematically extend to slightly larger values of $T/T_{\phi}$, as visible in Fig.~\ref{weak}. 
This could be attributed either to a systematic shift in the experimental determination of $T$ (e.g. arising in expansion imaging), which however increases with increasing $T/T_{\phi}$ ratio, or to the method by which the inputs to $T_{\phi}$ are extracted in the analysis.
In the rest of the paper we assume that this shift arises solely from the latter and attempt to further improve on the spanned range of $T/T_{\phi}$.

\subsection{\textbf{Identification of $T_{\phi}$} \label{Ident_Tphi} }

A given set of trap frequencies, temperatures and total atom numbers should fix the characteristic temperature $T_{\phi}$ to a particular value which is the same 
between theory and experiment; for this reason, in order to attempt to span precisely the same {\em range} of $T/T_\phi$ as in the experiments, we plot in the {\em inset} of Fig.~\ref{weak} our theoretically-obtained values for the reduced coherence length $L_c/L$ versus $T/T_\phi$ using here the {\em experimentally}-obtained values for $T_{\phi}$. This still reveals good agreement between our theoretical findings (blue filled circles) and the experimentally-extracted ones (hollow red circles), within the reported experimental error bars. 

We have also performed a separate analysis, based again on the SGPE but in which we instead match our numerically extracted {\em quasicondensate} number to the corresponding experimentally-extracted `condensate' number \cite{Jocelyn}, and measure the coherence length via $C^{(\rm 1,mod)}$.
These closely related measurements are 
shown in the inset of Fig.~\ref{weak} with green crosses; we find good overall agreement 
both with the experimental findings (hollow red circles) and with the previous analysis, based on matching the experimental total atom number instead (filled blue circles).

In order to further resolve the remaining discrepancy between theory and experiment in the regimes of $T/T_{\phi}$ analysed, in the next section we discuss an alternative method to extract $T_{\phi}$ from the SGPE simulations.

\subsubsection{\textbf{$T_{\phi}$ extracted from the phase distribution} }

In this section we investigate an alternative method of reproducing the experimental results of Ref.~\cite{Hugbart2005} from an SGPE analysis. 
Our approach is motivated from footnote 47 of \cite{Hugbart2005}, stating that $T_{\phi}$ can be obtained from the relation $L_{\phi}/L=T_{\phi}/T$ \cite{Petrov2001} with $L_{\phi}$ identified as the characteristic separation over which the phase fluctuates by $1$ radian at the trap centre.

In order to extract values for $T_{\phi}$ according to the relation described above, we
take here the approach of systematically analysing the phase distributions of the 
ensemble of stochastic fields ${\psi}$ at several distances from the trap centre.
Within the Thomas-Fermi radius, we find these distributions to be well fitted with Gaussian functions, whose standard deviation increases with increasing distance from the trap centre due to the enhanced role of thermal fluctuations (see Fig.~\ref{phase_histogram} in Appendix \ref{App} and related work in \cite{Betz2010}).
In this analysis, $L_{\phi}$ can be identified as the distance from the centre where the standard deviation of the Gaussian fit to the phase distribution reaches a particular value. Our analysis indicates that when this (free) parameter takes the value of $\sigma =0.65$, optimum agreement with the experimental findings is obtained regarding the spanned range of $T/T_{\phi}$. 

\begin{figure}[h]
  \includegraphics[scale=0.3]{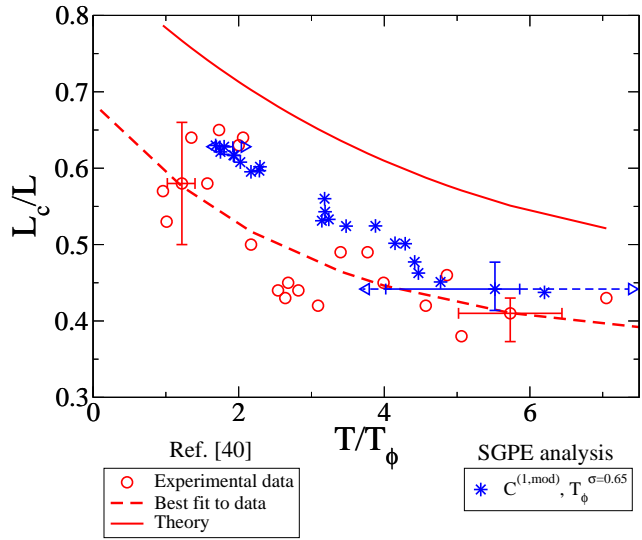}
   \caption {(Color online) Scaled coherence length vs. reduced temperature. Experimental data {\cite{Hugbart2005}} (hollow red circles) against numerical SGPE results extracted via $C^{{(\rm 1,mod)}}$ (blue stars); in the SGPE analysis temperatures are scaled to $T_{\phi}$ extracted from the phase distribution (see Appendix \ref{App}). The theory used in \cite{Hugbart2005} is also shown (solid line) together with the reported experimental best fit (dashed red line). Error bars account for $20\%$ variation in total atom number (solid blue line) and $8\%$ variation in the standard deviation (dashed blue line) (see Appendix \ref{App}). \label{phase_tphi_err_varyNandsigma} }
\end{figure}

The simulated results, with temperatures scaled to $T_{\phi}$ extracted as described above (blue stars), are shown in Fig.~\ref{phase_tphi_err_varyNandsigma}, alongside the experimental findings (hollow red circles). 
We also include here error bars for the previously considered indicative data points (as in Ref.~\cite{Hugbart2005}). In obtaining these we consider here two sources of error:
firstly, as in previous figures, a variation in
total atom number of $20\%$ \cite{Hugbart_PhD};
secondly, as the horizontal axis is also sensitive to the precise value of $\sigma$ chosen, the sensitivity to a variation in $\sigma$ of $\pm 0.05$ is shown by the additional dashed portions of the error bars.

\section{Conclusions}

An {\it ab initio} model capable of capturing the phase coherence properties of highly-elongated, weakly-interacting 
Bose gases is of fundamental importance for future applications such as atom-interferometers and atom lasers. In this work we have analysed one such model, the stochastic 
Gross-Pitaevskii equation (SGPE), with novel 
and self-consistent modifications in order to take account of quasi-one-dimensional effects. 
Comparing to the experiment of Richard {\em et al.}, Ref.~\cite{Richard2003}, which considered the regime 
of relatively strong phase fluctuations $T/T_{\phi}>6$,
we found excellent agreement between experiment and the SGPE theory, based solely on using 
the experimental parameters (i.e. trap configurations, atom number, temperatures) as inputs for the theory.

We further compared the SGPE theory to the findings of Hugbart {\em et al.}, Ref.~\cite{Hugbart2005} for the opposite low temperature regime $T \lesssim {\rm few}\hspace{0.1cm} T_{\phi}$, which is more challenging to probe in experiments.
While the SGPE analysis gave a temperature dependence of the coherence length in quantitative agreement with the experimental trend (within error bars), 
undertaking a point-by-point analysis of the experimental data was found to span a slightly narrower range of $T/T_{\phi}$ than the experimentally-reported curve, indicating a systematic deviation.
We argued that this discrepancy may arise as a result of the different identifications of $T_{\phi}$ between theory and experiment, possibly due to the different means of processing the `raw' experimental data and stochastic numerical results, e.g. due to differences in extracting the quasicondensate atom number which then feeds into the expression for $T_{\phi}$.

We have partially examined this issue by using instead a phase sensitive means of
extracting $T_{\phi}$ from characterisation of the ensemble phase distribution in SGPE simulations.
In particular, motivated by footnote 47 of Ref.~\cite{Hugbart2005}, we identified $T_{\phi}$ through the relation 
$T_{\phi} = T(L_{\phi}/R_{\rm TF}(T))$ where $L_{\phi}$ was chosen as the characteristic separation from the trap centre 
at which the phase distribution is fitted by a Gaussian  
with a particular value for the standard deviation.
In our treatment however this value is a free parameter chosen here so as to match the experimental range of $T/T_{\phi}$.

Finally, as the measurements performed in \cite{Hugbart2005} rely on an interferometric technique, this opens up a way to indirectly `extract' the Penrose-Onsager mode in a quasi-one-dimensional experiment solely from knowledge of density and the two lowest equal-time correlation functions, with the Penrose-Onsager condensate mode density emerging as the phase- {\em and} density-fluctuation suppressed part of the density via
$n_{\rm PO}(z) \approx n(z) \, \sqrt{2 - g^{(2)}(z)} \, g^{(1)}(0,z)$
\cite{AlKhawaja2002,Cockburn2011b}.

\section{Acknowledgments}
We acknowledge the group of Alain Aspect (in particular: Jocelyn Retter for providing us their data in 2005 and Mathilde Hugbart for recent clarifications), Fabrice Gerbier and Joseph Thywissen for comments specific to this manuscript and Mike Garrett for discussions. 
Funding was provided by the UK EPSRC (Grant No. EP/F055935/1).

\appendix
\section{Analysis of the phase distribution \label{App}}

The numerical procedure implementing the SGPE model makes use of the Condor system \cite{condor}, which is a mechanism to run parallel simulations, on different machines. 
\begin{figure}[h]
    \includegraphics[scale=0.3]{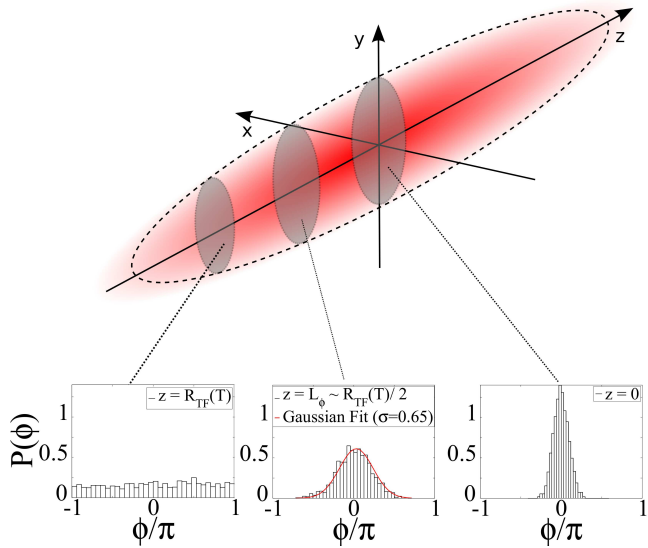}
  \caption {Distribution of the phase of $a_c^*\psi(z)/|a_c|$ (where $a_c$ is
the amplitude of the Penrose-Onsager mode), for a chosen temperature $T \approx T_{\phi}$ (left point with error bars in 
Figs.~\ref{strongandweak}-\ref{phase_tphi_err_varyNandsigma}), 
at different locations from the trap centre. Shown are the distributions at the condensate edge (left plot), at the point $z=L_{\phi}$ (central plot) where the distribution is fitted by a Gaussian (illustrated by the red solid curve) of standard deviation $\sigma=0.65$ (which for the particular numerical point considered occurs at $z \simeq 0.5 \, R_{\rm TF}(T)$) and at the trap centre $z=0$ (right plot). The distributions are
centered and normalised, and the phase $\phi$ is scaled to $\pi$.
\label{phase_histogram} }
\end{figure}
Using this method we generate a large number of stochastic realizations of the wavefunction $\psi$ (typically $\sim$ 1000 in a very short time); in this way we can then extract physical observables by performing the average over the ensemble of stochastic trajectories. 
In each numerical realization, which is somewhat analogous to an individual experimental run, the phase of the wavefunction $\psi$ takes on a certain value at each position, leading to a distribution across the ensemble; it is then interesting to investigate how the distribution of the several realizations of the phase changes with the distance from the trap centre, at a definite value of $T/T_{\phi}$.

In Fig.~\ref{phase_histogram} 
we report phase histograms of the stochastic field $\psi(z)$, locked to the phase of the Penrose-Onsager (condensate) mode $\phi_{\rm c}$ \citep{wright2011,Cockburn2011b}, at three values of the distance from the trap centre, for the experimental data point with $T \approx T_{\phi}$ (left point with error bar in Figs.~\ref{strongandweak}-\ref{phase_tphi_err_varyNandsigma}).
To be more specific, we plot the phase of $a_c^*\psi(z)/|a_c|$, where $a_c$ is the amplitude of the Penrose-Onsager mode, given by $a_c=\Delta z \sum_{z_{i}} \phi_c^*(z_i) \psi(z_i)$, where $\Delta z$ is the grid spacing
(see \citep{wright2011,Cockburn2011b} for further details and implementation).

The broadness of the generated distributions is an indication of the amount of coherence at a specific spatial point in the system: we expect the distribution to become broader with increasing distance from the trap centre, and be almost flat at the edge, where the system becomes purely thermal (Fig.~\ref{phase_histogram}).
We define $L_{\phi}$ to be the distance at which the phase histogram is well fitted by a Gaussian with a particular value of the standard deviation $\sigma$.
Following such a procedure, we can then define a phase fluctuation temperature $T_{\phi}$ from the relation $T_{\phi} \approx T(L_{\phi}/R_{\rm TF}(T))$ \cite{Petrov2000,Petrov2001}.
The standard deviation is here a free parameter, and we find that the value of $\sigma=0.65 \pm 0.05$ provides very good matching to the experimental data, as shown in Fig.~\ref{phase_tphi_err_varyNandsigma} (with corresponding range $z= L_{\phi} \sim (0.15-0.6) \, R_{\rm TF}(T)$ for this chosen value of $\sigma$).   


\bibliography{twa-sgpe2}

\begin{thebibliography}{100}
\bibitem{reconcile_tphi}%
  \BibitemOpen
  \bibinfo {howpublished} {The two expressions are reconciled through the
  relation $n_{\rm qc}(0)=15 N_{\rm qc}/16 L$ derived by integrating the
  density profile of a 3D Bose gas confined in an elongated cylindrical
  harmonic trap over the radial degrees of freedom within the Thomas-Fermi
  approximation.}%
  \bibAnnoteFile{Stop}{reconcile_tphi}%
\end{thebibliography}

\end{document}